\documentclass[conference]{IEEEtran}
\IEEEoverridecommandlockouts
\usepackage{amsmath,amssymb,amsfonts}
\usepackage{algorithmic}
\usepackage{graphicx}
\usepackage{textcomp}
\usepackage{xcolor}
\usepackage{makecell}
\usepackage{dirtytalk}
\usepackage{comment}
\usepackage{listingsutf8}
\usepackage{tabularx}
\usepackage{listings}

\lstdefinelanguage{docker}{
  keywords={FROM, RUN, COPY, ADD, ENTRYPOINT, CMD,  ENV, ARG, WORKDIR, EXPOSE, LABEL, USER, VOLUME, STOPSIGNAL, ONBUILD, MAINTAINER},
  keywordstyle=\color{blue}\bfseries,
  identifierstyle=\color{black},
  sensitive=false,
  comment=[l]{\#},
  commentstyle=\color{purple}\ttfamily,
  stringstyle=\color{red}\ttfamily,
  morestring=[b]',
  morestring=[b]"
}

\lstdefinelanguage{docker-compose}{
  keywords={image, environment, ports, container_name, ports, volumes, links, privileged},
  keywordstyle=\color{blue}\bfseries,
  identifierstyle=\color{black},
  sensitive=false,
  comment=[l]{\#},
  commentstyle=\color{purple}\ttfamily,
  stringstyle=\color{red}\ttfamily,
  morestring=[b]',
  morestring=[b]"
}
\lstdefinelanguage{docker-compose-2}{
  keywords={version, volumes, services},
  keywordstyle=\color{blue}\bfseries,
  keywords=[2]{image, environment, ports, container_name, ports, links, build},
  keywordstyle=[2]\color{olive}\bfseries,
  identifierstyle=\color{black},
  sensitive=false,
  comment=[l]{\#},
  commentstyle=\color{purple}\ttfamily,
  stringstyle=\color{red}\ttfamily,
  morestring=[b]',
  morestring=[b]"
}

\lstdefinelanguage{javascript}{
  morekeywords={typeof, new, true, false, catch, function, return, null, catch, switch, var, if, in, while, do, else, case, break},
  morecomment=[s]{/*}{*/},
  morecomment=[l]//,
  morestring=[b]",
  morestring=[b]'
}

\usepackage[sorting=none]{biblatex} %

\def\BibTeX{{\rm B\kern-.05em{\sc i\kern-.025em b}\kern-.08em
    T\kern-.1667em\lower.7ex\hbox{E}\kern-.125emX}}
\usepackage[T1]{fontenc}
\usepackage{verbatim}
\newcolumntype{Y}{>{\centering\arraybackslash}X}

\newcolumntype{P}[1]{>{\centering\arraybackslash}p{#1}}
\newcolumntype{M}[1]{>{\centering\arraybackslash}m{#1}}

\begin{document}
\title{Dockerized Android: a container-based platform to build mobile Android scenarios for Cyber Ranges
{
\thanks{}
}}

\author{
\IEEEauthorblockN{Daniele Capone}
\IEEEauthorblockA{\textit{SecSI srl} \\
Napoli, Italy \\
daniele.capone@secsi.io}
\and

\IEEEauthorblockN{Francesco Caturano}
\IEEEauthorblockA{\textit{Dept. of Electrical Engineering and Information Technology} \\
\textit{University of Napoli Federico II}\\
Napoli, Italy \\
francesco.caturano@unina.it}

\and

\IEEEauthorblockN{Angelo Delicato}
\IEEEauthorblockA{\textit{SecSI srl} \\
Napoli, Italy \\
angelo.delicato@secsi.io}

\and

\IEEEauthorblockN{Gaetano Perrone}
\IEEEauthorblockA{\textit{Dept. of Electrical Engineering and Information Technology} \\
\textit{University of Napoli Federico II}\\
Napoli, Italy \\
gaetano.perrone@unina.it}

\and

\IEEEauthorblockN{Simon Pietro Romano}
\IEEEauthorblockA{\textit{Dept. of Electrical Engineering and Information Technology} \\
\textit{University of Napoli Federico II}\\
Napoli, Italy \\
spromano@unina.it}}

\maketitle
\begin{abstract}
The best way to train people about security is through Cyber Ranges, i.e., the virtual platform used by cyber-security experts to learn new skills and attack vectors. In order to realize such virtual scenarios, container-based virtualization is commonly adopted, as it provides several benefits in terms of performance, resource usage, and portability. Unfortunately, the current generation of Cyber Ranges does not consider mobile devices, which nowadays are ubiquitous in our daily lives. Such devices do often represent the very first entry point for hackers into target networks. It is thus important to make available tools allowing to emulate mobile devices in a safe environment without incurring the risk of causing any damage in the real world. This work aims to propose Dockerized Android, i.e., a framework that addresses the problem of realizing vulnerable environments for mobile devices in the next generation of Cyber Ranges. We show the platform's design and implementation and show how it is possible to use the implemented features to realize complex virtual mobile kill-chains scenarios.
\end{abstract}

\begin{IEEEkeywords}
Container-based architectures, Mobile Cyber Ranges, Mobile Android Hacking, Security.
\end{IEEEkeywords}

\section{Introduction}
Cyberthreats are real, and they will always be. There will never be 100\% secure software. What an organization or a country can do is avoid a security incident from a cyber threat and, if that happens despite all the efforts, minimize the consequences. In order to do so, the organization/country needs highly trained personnel. Currently, one of the most effective ways to train people about security is through \textit{Cyber Ranges}, i.e., virtual platforms that allow companies and security enthusiasts to learn cyber-security skills in a safe and controlled environment. These platforms can be instrumented for several purposes, even if they are commonly used to create challenges to train security teams. Such challenges are formed by two teams: the ``Red Team'', i.e., security experts that help companies to discover vulnerabilities by mimicking a real attacker, and the ``Blue Team'', i.e., security experts that try to protect the vulnerable infrastructures from the Red Team. Although it is proven that cyber ranges are essential to improve the effectiveness of security training~\cite{8190713}, the problem with current Cyber range generations is that they usually do not consider the existence of mobile devices. This is a strong limitation, as nowadays, mobile devices are commonly adopted by companies and represent a severe security risk. Proofpoint security researchers show that the number of attacks increased by 500\% at the beginning of 2022 \cite{proof}. A mobile device can browse the Internet, connect to an FTP server, and access all data. Therefore, it could be easily the entrance point for hackers to introduce themselves into the target network. Mayrhofer et al. (2021) \cite{Mayrhofer2021} provide an Android threat model that covers several mobile threats' scenarios. 
The threats summarized in Table~\ref{tab:threats} can be physical and/or proximal (P), based on network attacks (N), caused by applications' abuse (A), or by untrusted data processing. 
\begin{table}[htb!]
\caption{Threat Model for Mobile Devices}
\label{tab:threats}
\begin{tabularx}{0.5\textwidth}{|c|Y|} 
 \hline
\textbf{TID}             & \textbf{Description}                                                           \\ \hline
T.P1                     & Physical attack through remote vectors, e.g., cellular, Wi-Fi, Bluetooth, GPS, NFC, and FM).                                     \\ \hline
T.P2 & Mobile shutdown through physical access.                                                                                        \\ \hline
T.P3                     & Device stolen by the attacker.                                                                                           \\ \hline
T.P4                     & Authorization abuse through access to the device.                                                                              \\ \hline
T.C1                     & Passive eavesdropping through device network identifiers as Media Access Control.                                        \\ \hline
T.C2                     & Man In The Middle attacks.                                                                                                      \\ \hline
T.A1                     & Abusing APIs supported by the OS with malicious intent, e.g., spyware.                                                         \\ \hline
T.A2                     & Abusing APIs supported by other apps installed on the device.                                                                 \\ \hline
T.A3                     & Untrusted code from the web (i.e., JavaScript) is executed without explicit consent.                                            \\ \hline
T.A4                     & Mimicking system or other app user interfaces to confuse users.                                                               \\ \hline
T.A5                     & Reading content from system or other app user interfaces.                                                                       \\ \hline
T.A6 & Injecting input events into the mobile OS system or through other app user interface.                                                                 \\ \hline
T.D1 & Perform spanning attacks against the victim's device through mobile calls, SMS, or emails.                                      \\ \hline
T.D2 &  Exploiting code that processes untrusted content in the OS or apps through remote or local vulnerabilities. \\ \hline
\end{tabularx}
\end{table}

Therefore, security training for these scenarios is crucial, and cyber-ranges can give a great contribution to such a purpose.

This paper describes the design and implementation of Dockerized Android, i.e., a platform that allows emulating Android devices in a container-based environment that represents a fully realistic yet inherently safe place to study and train.

The paper is organized in six sections. 
Section~\ref{sec:related} discusses related works associated, respectively, with the introduction of mobile components in Cyber Ranges and the container-based emulation of Android devices. Section~\ref{sec:design} proposes a bird's eye view of the proposed architecture by providing a list of requirements to fulfill, as well as some possible usage scenarios. In Section~\ref{sec:implementation} dig deeper into the details associated with its implementation through Docker. %
Section~\ref{sec:evaluation} provides a qualitative evaluation of what has been done with the help of a few practical usage examples. Finally, Section~\ref{sec:conclusions} discusses how the proposed architecture naturally lends itself to the study of advanced mobile hacking scenarios like the so-called ``mobile cyber kill chain''. It also highlights interesting directions for our future work in the field.

\section{Related Works}
\label{sec:related}
Our platform leverages the benefits of container-based virtualization, an approach extensively adopted to create cyber-ranges. 
Several authors explore the applicability of container-based solutions to the cyber-range domain. Nataka et al. (2020)~\cite{nakata} confirm the performance benefits of using container-based virtualization techniques to reproduce cyber ranges.
In another work, Nakata and Otsuka (2020)~\cite{Nakata2020} evaluate the vulnerability coverage of container-based solutions and estimate that $99.3$\% of vulnerabilities are reproducible.
Despite these results, we think that container-based virtualization has several limitations in terms of vulnerability reproduction. This is discussed in detail in~\cite{flags}. 

Even though there is rich literature about cyber ranges~\cite{Yamin2020}, few researchers have addressed the problem of enhancing them with support for mobile scenarios.
One of the most recent works that try to fill this gap is authored by Russo et al. (2020)~\cite{next-generation-cyber-ranges}. The authors give a full overview of the current scenario about both cyber ranges and security training before going into a  more detailed description of the work that needs to be done to create \textit{Next-Generation Cyber Ranges}. The work introduces an attack scenario inspired by Pierini and Trotta (2017)~\cite{from-apk-to-golden-ticket} that leverages the so-called ``Kerberos Golden Ticket'' to obtain administrative privileges. In the proposed scenario, the attacker has no a priori knowledge and time-limited access to the target system. The attacker must proceed through several stages to hack the system, and the last three steps involve the use of mobile techniques.
The proposed approach emphasizes the integration tasks needed to build cyber ranges with mobile system support. One of the most significant issues is licensing problems and virtual hardware compatibility. There are indeed several solutions to that problem.
\textit{Genymotion}~\cite{genymotion} is a commercial solution that allows the execution of Android virtual devices in three different ways:
\begin{itemize}
    \item \emph{Desktop}: allows running Android Virtual Devices on one's own desktop computer (like the official Android emulator);
    \item \emph{SaaS}: allows running Android virtual devices on Genymobile Servers (mostly for test automation);
    \item \emph{PaaS Images}: virtual Android images for Cloud providers.
\end{itemize}

\textit{Corellium}~\cite{corellium} is yet another commercial solution that allows to run ARM Virtualized Devices in the Cloud. Such a solution provides different features for different use cases, e.g., security research, app streaming, and device modeling.
To solve the licensing issues and take advantage of container-based capabilities, we leverage the official Android Emulation platform~\cite{androidemu} to build our mobile cyber-range environment. This choice is justified by the public availability of several solutions allowing to have a Docker Image with a pre-installed Android SDK (and an Emulator as well):
\begin{itemize}
    \item \textit{thyrlian/AndroidSDK}: a fully-fledged Android SDK Docker image~\cite{thyrlian-android-sdk};
    \item \textit{budtmo/docker-android}: an `Android in Docker' solution with support for \emph{noVNC}, as well as video recording capabilities~\cite{budtmo-docker-android};
    \item \textit{bitrise-io/android}: a customized Android Docker image~\cite{bitrise-android}.
\end{itemize}

These solutions are based on the SDK and have been realized for development and testing purposes. In particular, they leverage the benefit of container-based virtualization in terms of performance, scalability, and portability but are not able to fully satisfy the security research and Cyber range needs. In particular, these solutions do not offer features that help the realization of scenarios that reproduce the threats described in Table~\ref{tab:threats}. However, they are a good starting point to work on for solving the Mobile Virtualization problem.

\section{Dockerized Android: design}
\label{sec:design}

Nowadays, it is imperative to integrate mobile systems into a Cyber Range due to the rising importance and ubiquitous usage of this kind of device. In~\cite{next-generation-cyber-ranges} the difficulty of virtualizing mobile systems is highlighted as the main problem associated with this kind of integration.

The main purpose of our work is actually the implementation of a framework that allows the integration of a mobile system in a container-based environment. 
We have focused our efforts on Android for several reasons:
\begin{itemize}
    \item it is open-source, so it is possible to easily reproduce the target environment;
    \item it runs on nearly every device;
    \item there are many external tools that can easily be integrated to provide enhanced features.
\end{itemize}

A further reason for focusing on Android is related to compatibility issues. In the implemented system, the Android Emulator is used, and it is both open source and compatible with all host operating systems (Linux, Windows, and OS X). The iOS Simulator, on the other hand, is a closed platform that can only run on OS X, thus making it impossible to test and work on without a machine running OS X.
We have opted for a container-based solution to take advantage of the associated performance improvements~\cite{nakata}, as well as for the ever-increasing usage of this kind of virtualization over the traditional virtual machine approach in current applications.
In particular, the resulting implementation should allow to:
\begin{itemize}
    \item run an Android virtualized device;
    \item integrate a physical device;
    \item control this device through a web browser;
    \item control this device through the \textit{ADB} (Android Debug Bridge) tool;
    \item configure the main virtualized components;
    \item install applications;
    \item collect data;
    \item define networking info;
    \item configure usage options;
    \item integrate other tools.
\end{itemize}

We formalize these requirements as ten functional (\textit{F}) requirements and five non-functional \textit{NF} requirements.

\emph{[F01] - Android virtual device execution.} The resulting system has to allow the execution of an Android virtualized device in a container-based environment. Before the execution, the user should also be able to select a few significant options (like, e.g., the specific Android OS version).

\emph{[F02] - Web browser device management.} The user has to be able to access the device through a web browser using an ad hoc front-end. This front-end has to provide all the features through a clean and simple UI (i.e., the device has to be usable through such a UI).

\emph{[F03] - ADB device management} The user has to be able to access the device through the Android Debug Bridge either directly from the above-mentioned front-end UI or by accessing the container of the device itself. 

\emph{[F04] - Virtualized components configuration.} The resulting system has to allow the user to configure all of the virtualized components of the emulated system. As an example, the user should be able to configure the GPS location of the device or use its SMS features. For obvious reasons, some hardware components cannot be fully virtualized. For example, the \textit{Android Emulator} virtualizes the camera, but it is nearly impossible to use it in an attack chain. Some other components, like the microphone, could be virtualized and used in a simulated attack chain. A proper distinction is provided in Section~\ref{sec:evaluation}.

\emph{[F05] - Application management.} The resulting system has to allow the user to manage the applications on the device. Namely, the user has to be able to install applications from files, as well as to install applications even from the \textit{Play Store} (this last feature requires a Google Account, and for this reason, it is not mandatory).

\emph{[F06] - Data collection.}
The resulting system has to allow the user to collect data and record the sessions of the platform. This feature is particularly useful for cyber-range administrators interested in reviewing the actions performed by the users.

\emph{[F07] - Networking configuration.} The resulting system has to allow the user to configure the network before executing the container. For example,  the user has to be able to configure the IP addresses of the devices. Another advanced feature is the ability to expose specific ports of the virtualized device, which are used to interact with specific device services. 

\emph{[F08] - Features configuration.} All of the features described in the previous requirements must be configurable, meaning that before the execution of the container, the user can set some options or even decide the activation of a particular feature to customize the environment entirely.

\emph{[F09] - Third-party tools integration.} The platform has to allow third-party tools integration with minimum effort, meaning that it should expose the proper interfaces in order to communicate with a third-party container that can be used as a tool. For example, the \textit{Mobile Security Framework} (MobSF)~\cite{mobsf} could be integrated and used in a particular configuration that allows the user to test the security of an app using the virtualized system.

\emph{[F10] - Physical device integration.} The system has to be able to integrate an Android physical device fully. Even when a physical device is connected, all of the features described previously must be usable in the same way.

\emph{[NF01] - Management UI scalability.} The front-end UI described in the \emph{requirement F02} has to be adaptive, meaning that it can be used with different screen resolutions. 

\emph{[NF02] - Image size shrinking.} The resulting system (that could be either a single Docker image or multiple ones) has to be developed by trying to reduce as much as possible the dimensions of the resulting image to save space on the device that actually runs the system. In order to optimize the environment, we followed all the best practices suggested by the Docker Community~\cite{dockerfiles} to keep the container's size as low as possible.

\emph{[NF03] - Separation of concerns.} The Docker Community switched to the 'single concern' principle~\cite{dockerfiles}, meaning that a single Docker image can run multiple processes if related to the same concern. The framework should keep this principle in mind when implementing its own images.

\emph{[NF04] - Host cross-platform compatibility.} The platform must be usable on different operating systems (i.e., Linux, macOS, Windows). 

\emph{[NF05] - Device type transparency.} The resulting system has to also allow for the execution of a physical Android device transparently. The physical device should get connected via either a wired or a wireless communication channel. 

Table \ref{tab:design-requirements} summarizes the above features; it is also used as a reference model in section~\ref{sec:evaluation}:

\begin{table}[htb!]
    \caption{Design requirements}
    \label{tab:design-requirements}

    \centering
    \begin{tabular}{|c|c|c|}
        \hline
        \textbf{ID} & \textbf{Type} & \textbf{Description} \\
        \hline
        F01 & Functional & Android Virtual Device execution \\
        \hline
        F02 & Functional & Browser management \\
        \hline
        F03 & Functional & ADB device management \\
        \hline
        F04 & Functional & Virtualized components configuration \\
        \hline
        F05 & Functional & Application management \\
        \hline
        F06 & Functional & Data collection \\
        \hline
        F07 & Functional & Network Configuration \\
        \hline
        F08 & Functional & Features Configuration \\
        \hline
        F09 & Functional & Third-party tools integration \\
        \hline
        F10 & Functional & Physical device integration \\
        \hline
        NF01 & Non-Functional & Management UI scalability \\
        \hline
        NF02 & Non-Functional & Image size shrinking \\
        \hline
        NF03 & Non-Functional & Separation of concerns \\
        \hline 
        NF04 & Non-Functional & Host Cross-Platform compatibility \\
        \hline
        NF05 & Non-Functional & Device type transparency \\
        \hline
    \end{tabular}
\end{table}

\section{Dockerized Android Architecture}
\label{sec:implementation}
This section shows the architecture of Dockerized Android.
Fig.~\ref{fig:high} provides a high-level view of the \emph{Dockerized Android} platform that allows identifying the key components it comprises: 

\begin{figure}[htb!]
\centering
\includegraphics[width=0.45\textwidth]{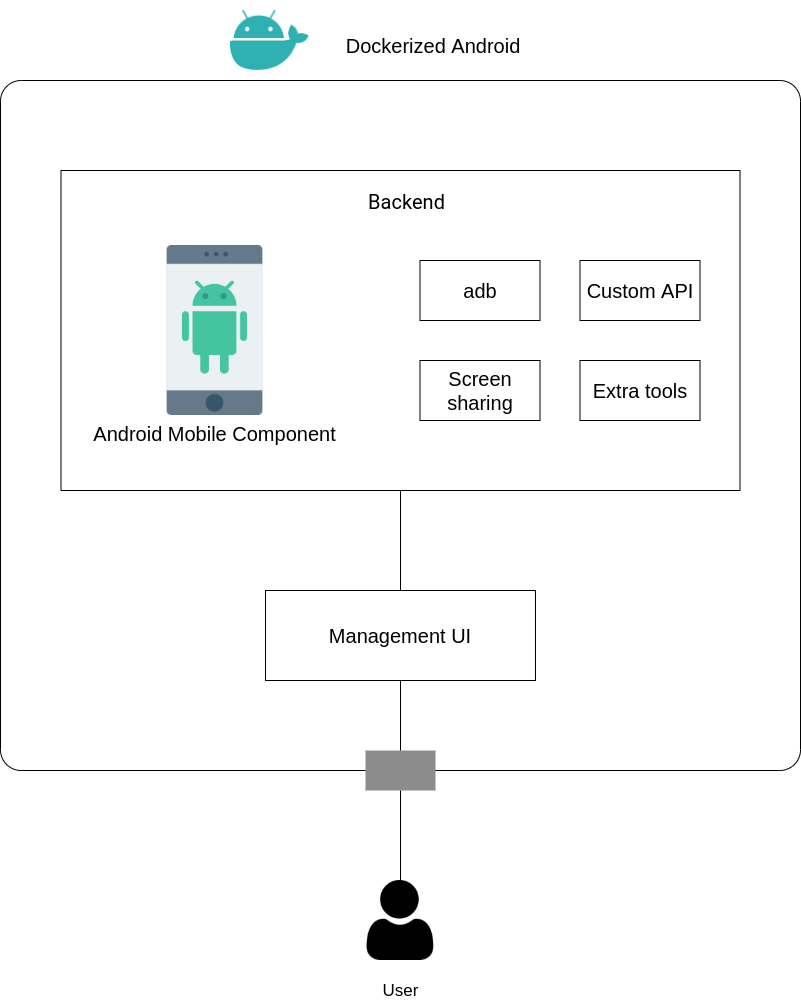}
\caption{Overall Architecture, resources from flaticon.com \cite{flaticon}}
\label{fig:high}
\end{figure}
\noindent
\begin{itemize}
    \item \emph{Android Mobile Component}: is the Android system accessible from the other components. It can be indifferently an emulated device or a real one connected to the machine hosting the container (through a cable or wireless). This component has to provide all of the features related to the Android OS itself, but also the hardware components that can be simulated (like the GPS or the microphone through the integration of an external device);
    \item \emph{adb}: is the well-known Android Debug Bridge allowing to control an Android system (either an emulated one or a real one) through a command-line interface. This component allows having a shell on the device, configure options, install applications, etc. It has to be externally accessible to provide a higher level of configuration and increase the end-user experience as well;
    \item \emph{Screen Sharing}: the goal of this component is to provide a server that can be used by another component in order to give the user a simpler way to access and control the mobile device;
    \item \emph{Custom API}: as several features are integrated into the system, this component provides access to the external tools with a uniform interface that hides the underlying cumbersome integration mechanisms;
    \item \emph{Management UI}: is the application front-end that can be accessed by the user to control the mobile device and enable the other features provided by the platform;
    The UI uses the VNC (Virtual Network Computing) component to let the user control the device through a web browser, the abd component to provide a shell to interact with the device in a more advanced way, and the Custom API to provide access to other features effortless;
    \item \emph{Extra Tools}: this component encloses the external tools used to add further features to the platform.
\end{itemize}
Fig.~\ref{fig:detailed-arch} shows a in-depth view of the modules implemented in the framework:
\begin{figure}[htb!]
\centering
\includegraphics[width=0.45\textwidth]{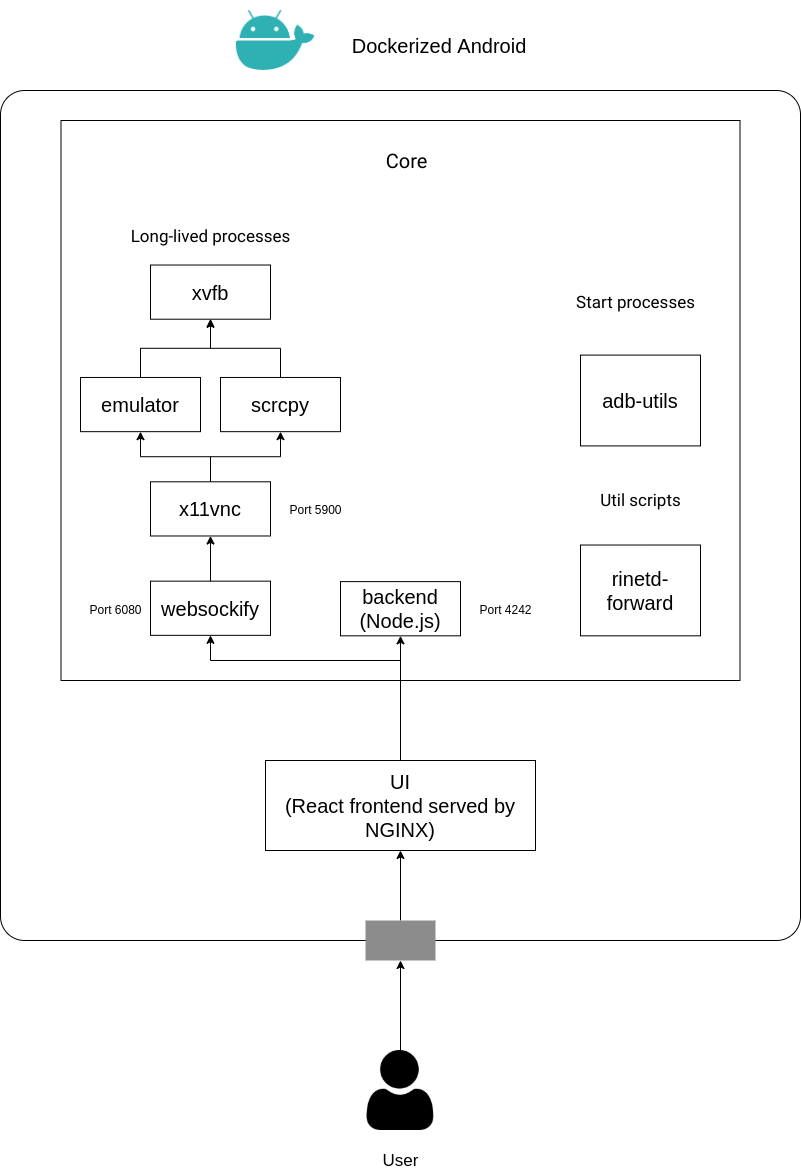}
\caption{Dockerized Android detailed architecture}
\label{fig:detailed-arch}
\end{figure}
The architecture can be divided into two parts:
\begin{itemize}
    \item \textit{Core}: executes all the processes needed to run an Android Component (Emulated or Real) inside a Docker container, also exposing some features to the outside.
    \item \textit{UI}: gives the user a UI to access the Core component in a simpler way through a web browser. 
\end{itemize}
The Core component is composed of several modules that execute ``long-lived processes'' (i.e., processes that offer several functionalities along with all the Dockerized Android execution), ``start processes'' (i.e., processes executed during the framework's bootstrap phase), and ``utility scripts''. 
The internal modules implement several features:
\begin{itemize}
        \item \textit{xvdb}, \textit{srcpy}, and \textit{x11vnc} modules simulate a real display and synchronize the device's monitor with a virtual X server in the host system (i.e., the computer system that executes Dockerized Android).
        \item The \textit{websockify module} converts the VNC communication protocol used by x11vnc to Web Socket network protocol; in this way, it is possible to view the mobile's screen through a Web browser.
        \item The \textit{emulator component} manages the Android Virtual Device that emulates the mobile device when an emulated device is used.
        \item The \textit{backend module} written in Node.js implements an extendible interface that allows adding cyber-range focused features, such as the dispatch of a malicious SMS or e-mail to simulate phishing attacks. 
        \item The \textit{adb-utils} module implements several utility scripts that use the ADB tool for implementing other functionality, such as the installation of vulnerable Android applications or instruments the mobile device with mobile security toolkits, such as Frida \cite{frida}. 
        \item The \textit{rinetd-forward} module manages the emulator's port forwarding by enabling the network communication between all the components. 
    \end{itemize}

The UI component is composed of a \textit{React Frontend} and is served through an \textit{NGINX server}. The UI is developed using the most recent features provided by the React framework, like Hooks and Context, in order to follow the principle of strong cohesion and loose coupling. The UI provides a simple way to use all the features exposed by the backend and also adds the ability to display and control the device. The user has to manually insert the address of the Core component and the corresponding ports (the port exposed by the backend and the port exposed by websockify.
\begin{figure}[htb!]
\centering
\includegraphics[width=0.48\textwidth]{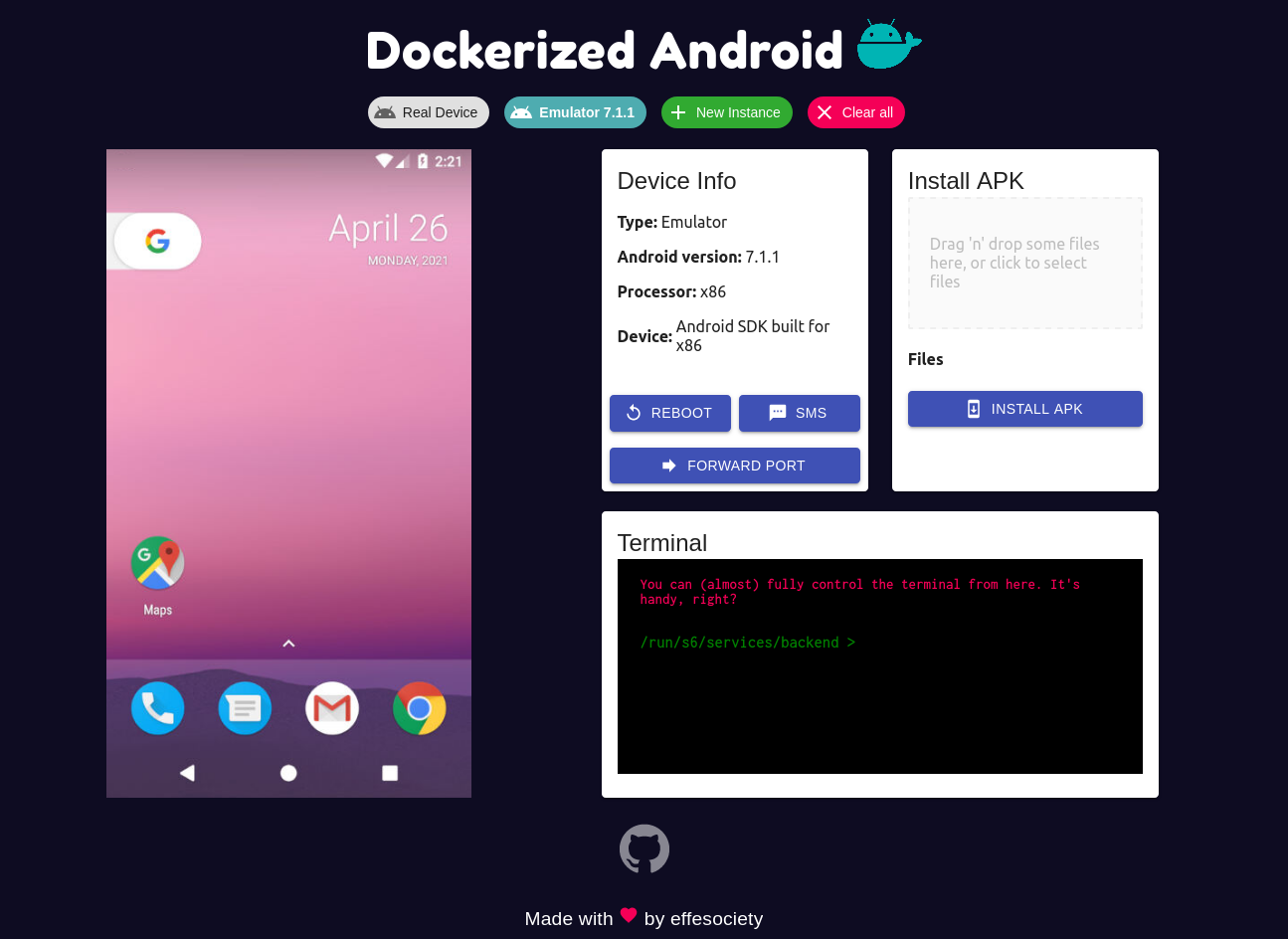}
\caption{UI Dockerized Android}
\label{fig:features}
\end{figure}
\noindent
Fig.~\ref{fig:features} shows the most interesting Dockerized Android user interface view.

The implementation of the above-mentioned architecture is publicly accessible on GitHub~\cite{dockerized-android}.

\section{Evaluation}
\label{sec:evaluation}

This section assesses the \emph{Dockerized Android} platform by examining several aspects. Firstly, we emphasize the differences between the \textit{Core for Emulator} and \textit{Core for Real Device} components in terms of features and highlight compatibility with the three most used Operating Systems. Then, we provide practical usage examples of \emph{Dockerized Android} and discuss coverage of the requirements defined in Section~\ref{sec:design}.

\subsection{Differences between Core for Emulator and Core for Real Device}
Even if a significant effort has been put into creating a system that has the same features for both kinds of devices, there are limitations when emulation is used: 
\begin{itemize}
    \item \emph{SMS ADB send/reception feature}: in emulated devices, it is possible to automate the send and reception of SMS messages through the ADB software. Obviously, this is not natively possible for real devices. Therefore, the user must manually send and receive SMS messages to implement SMS attack scenarios. A solution to address this problem could be the realization of a custom Android application that could be installed on a real device and could be instrumented to send and receive messages automatically.
    \item \emph{Networking}: networking is quite different between the Emulator and the Real device flavors. In the emulator version, the AVD is created inside the Docker container, and therefore it shares the container's IP address. Instead, the real device is physically connected to the machine that runs the container and keeps its own IP address.
    \item \emph{Hardware virtualization}: for the hardware components, the situation is quite different, too: some hardware devices like the GPS and the microphone can be emulated. In particular, the GPS location of the device can be set through ADB, and the microphone of the host machine can be shared with the emulator. There are other hardware components that currently cannot be emulated, like, e.g. Bluetooth.

\end{itemize}

\subsection{Host evaluation for cross-platform compatibility}
The non-functional requirement \emph{NF04 (Cross-platform compatibility)} states that the resulting system should be usable from within any host OS. This refers to the OS of the machine that runs the Docker containers. Table~\ref{tab:comparison-os-compatibility} provides a summary of the compatibility with Linux, Windows, and OS X.

\begin{table}[htb]
    \caption{Host OS compatibility comparison}
    \label{tab:comparison-os-compatibility}

    \centering
    \begin{tabular}{|c|c|c|}
        \hline
        & \textbf{Core for Emulator} & \textbf{Core for Real Device} \\
        \hline
        \textbf{Linux} & Full Compatibility & Full Compatibility \\
        \hline
        \textbf{Windows} & Not supported (yet) & Full Compatibility \\
        \hline
        \textbf{OS X} & Not supported & Workaround \\
        \hline
    \end{tabular}
\end{table}
\noindent
The problem with Windows is that currently, the best way to use Docker is through the \textit{Windows Subsystem for Linux} (WSL) framework. Unfortunately, WSL does not support nested virtualization yet, and this feature is required to run the Android emulator inside a Docker container. However, the feature will be available in upcoming WSL releases. It might be possible to run the \textit{Core for Emulator} flavor on Windows by using a virtual machine, though losing all of the performance benefits associated with containerization.
A similar issue does exist with OS X, with which there is currently no way to run the \textit{Core for Emulator}. 
Besides, OS X does not allow sharing the USB device with a Docker container.
For this reason, the only ways to use the \textit{Core for Real Device} flavor are to either run ADB over Wi-Fi or connect to the host ADB from within the Docker container.

In the remainder of this section, we show the effectiveness of \emph{Dockerized Android} in reproducing security kill chains by using both the \textit{Core for Emulator} and \textit{Core for Real Device}.

\subsection{Security attack reproduction on the emulator}

We herein focus on a sample vulnerability scenario associated with CVE-2018-7661\footnote{\url{https://cve.mitre.org/cgi-bin/cvename.cgi?name=CVE-2018-7661}}. This CVE is related to the free version of the application ``Wi-Fi Baby Monitor''. This application has to be installed on two devices in order to act as a so-called \emph{baby monitor} (a radio system used to remotely listen to sounds emitted by an infant). 
As reported in the National Vulnerability Database, ``Wi-Fi Baby Monitor Free \& Lite'' before version $2.02.2$ allows remote attackers to obtain audio data via certain specific requests to TCP port numbers $8258$ and $8257$''.

\begin{table}[htb!]
    \caption{Requirements for Wi-Fi Baby Monitor}
    \centering
    \begin{tabular}{|c|c|}
        \hline
        \textbf{Device Type} & Real Device or Emulator \\
        \hline
        \textbf{Application} & Wi-Fi Baby Monitor \\
        \hline
        \textbf{APK Version} & Free \& Lite before 2.02.2  \\
        \hline
    \end{tabular}
\end{table} 
\noindent

The premium version of this application offers users the ability to specify a password to use in the pairing process. By monitoring the network traffic, it is possible to observe that:
\begin{itemize}
    \item the initial connection takes place on port $8257$;
    \item the same sequence is always sent to start the pairing process;
    \item at the end of the pairing process, a new connection is started on port $8258$. This port is used to transmit the audio data;
    \item after connecting to the port $8258$, the other connection on the port $8257$ is kept open and used as a heartbeat for the session;
    \item on the heartbeat connection, the client periodically sends the hexadecimal byte $0x01$ (about once per second);
\end{itemize}

The proof of concept that allows the attacker to obtain audio data is given in \cite{baby-monitor-poc}. This Proof of Concept (PoC) is easily reproducible on Dockerized Android through the realization of an infrastructure composed of three services: 

\begin{itemize}
    \item \emph{core-emulator}: an instance of the Core component with a pre-installed Baby Monitor app acting as the sender;
    \item \emph{ui}: the UI component to control what is going on;
    \item \emph{attacker}: a customized version of Kali Linux that automatically installs all the dependencies needed for the execution of the PoC.
\end{itemize}

This is also a perfect example to show the \emph{Port Forwarding} feature used to enable the communications.

\subsection{Security attack reproduction on the real device}

With the real device, we examine a further vulnerability, known as \emph{BlueBorne}. The term ``BlueBorne'' refers to multiple security vulnerabilities related to the implementation of Bluetooth. These vulnerabilities were discovered by a group of researchers from Armis Security, an IoT security company, in September $2017$. According to Armis, at the time of discovery, around $8.2$ billion devices were potentially affected by the BlueBorne attack vector, which affects the Bluetooth implementations in Android, iOS, Microsoft, and Linux, hence impacting almost all Bluetooth device types such as smartphones, laptops, and smartwatches. BlueBorne was analyzed in detail in a paper published on the $12^{th}$ of September $2017$ by Ben Seri and Gregor Vishnepolsk~\cite{technical-paper-blueborne}. Eight different vulnerabilities can be used as part of the attack vector.%

Regarding Android, all devices and versions (therefore versions older than Android Oreo, which was released in December $2017$) are affected by the above-mentioned vulnerabilities, except for devices that support BLE (Bluetooth Low Energy).
In general, two requirements should be satisfied to exploit the vulnerability: (i) the target device must have Bluetooth enabled; (ii) the attacker must be close enough to the target device. 
As the Bluetooth feature is not available in the Core Emulator, the kill-chain in question can only be reproduced on real devices.

\subsubsection{BlueBorne full reproduction on Dockerized Android}
In order to show the effectiveness of Dockerized Android, we developed a kill chain that exploits two Remote Code Execution (RCE) vulnerabilities that affect Android, i.e., CVE-2017-0781 and CVE-2017-0782. These vulnerabilities fall within the Bluetooth set vulnerability's set defined ``BlueBorne'' and discovered by a group of security researchers from Armis Security \cite{armis}.  

\begin{figure}[htb!]
\centering
\includegraphics[width=0.48\textwidth]{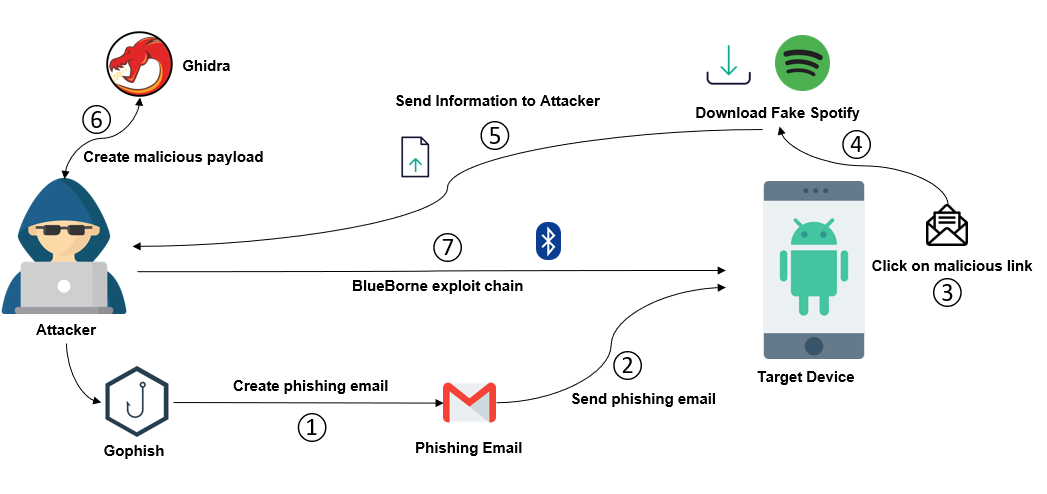}
\caption{Exploit Chain Overview}
\label{fig:echain}
\end{figure}

The diagram in Fig.~\ref{fig:echain} gives an  overview of the developed kill chain:
\begin{enumerate}
    \item The attacker creates a phishing email through Gophish, a phishing generator software. 
    \item The phishing email is sent to a victim's mailbox. 
    \item The victim reads the phishing email and erroneously clicks a malicious link contained in the email's body.
    \item The malicious link allows the attacker to trigger an attack that downloads and installs a fake application on the victim's mobile device. 
    \item The malicious information sends relevant mobile information to the attacker. This information is required for the exploitation of the two vulnerabilities.
    \item The attacker crafts a malicious payload to exploit the vulnerabilities. 
    \item The attacker sends the attack by exploiting the Bluetooth component's vulnerabilities and has remote access to the victim's device.
\end{enumerate}
The complex scenario covers several threats defined in Table~\ref{tab:threats}. Table \ref{tab:coverage} shows such threats and both the platform functionalities and components that allow the scenario reproduction. 
\begin{table}[htb!]
\caption{Threats, scenario's steps, features and components}
\label{tab:coverage}

\def\arraystretch{1.5}
\begin{tabularx}{0.5\textwidth}{|Y|Y|Y|Y|}
\hline
\textbf{Threat ID} & \textbf{Scenario step} & \textit{\textbf{Useful features}} &  \textit{\textbf{Dockerized Android components}} \\ \hline
T.D1               & 2,3                    & F01, F02, F08, F10                                            & xvdb, scrpy, x11vnc \\ \hline
T.A4               & 4                      & F02, F03   & websockify, UI, backend, adb-utils                                                  \\ \hline
T.A1, T.D2         & 5                      & F07    &   rinetd-forward                                                 \\ \hline
T.P1               & 7                      & F10      & backend                                                  \\ \hline
\end{tabularx}
\end{table}
The scenario requires complex network communications (F07) and involves the utilization of Bluetooth. For this reason, we have to use a physical device (F10). In the proposed scenario, we have to simulate the installation of the malicious application when the user receives the email. This can be done either manually (F02) or by implementing utility ADB scripts (F03).  
In order to reproduce the scenario, additional elements are needed: 
\begin{itemize}
    \item \emph{Gophish}: a webapp that allows to craft and send phishing emails, for which a Docker version already exists.
    \item \emph{Ghidra}: an application created by the National Security Agency (NSA) for reverse engineering purposes. In this context, it is used to get some useful information about the target device. This application is used on the host machine without Docker.
    \item \emph{Fake Spotify}: a seemingly benign application that pretends to provide the user with a free version of the well-known Spotify Premium app, but rather sends to the attacker's server exfiltrated files that are reverse-engineered on Ghidra. Also, this app was created without the usage of Docker.
\end{itemize}

Listing~\ref{lst:compose} shows how it is possible to simulate a container-based virtualized scenario by defining a docker-compose file, i.e., a declarative file interpreted by the Docker engine to generate the virtual infrastructure.

\begin{lstlisting}[language=C, breaklines=true, frame=single,caption=docker-compose.yaml for the BlueBorne kill chain,label=lst:compose, ,captionpos=b]
version: "3.8"
services:
  core-real:
    image: secsi/dockerized-android-core-real-device
    privileged: true
    networks:
      blueborne-net:
        ipv4_address: 10.5.0.2
  ui:
    image: secsi/dockerized-android-ui
    ports:
      - "8080:80"
    networks:
      blueborne-net:
        ipv4_address: 10.5.0.3
  attacker_phishing:
    image: gophish/gophish
    ports:
      - "3333:3333"
      - "8081:8080"
    volumes:
      - ./phishing:/home/phishing
    networks:
      blueborne-net:
        ipv4_address: 10.5.0.4
  attacker_blueborne:
    image: secsi/kali-rolling-with-dependencies:latest
    tty: true
    volumes:
      - ./exploit:/home/exploit
      - ./dependencies-blueborne:/home/dependencies
    environment:
      - SH_DEPENDENCIES_FILE_PATH=/home/dependencies/
      dependencies.sh
    privileged: true
    network_mode: "host"
  attacker_web_server:
    image: secsi/kali-rolling-with-dependencies:latest
    tty: true
    ports:
      - "8000:8000"
    volumes:
      - ./webserver:/home/webserver
      - ./dependencies-webserver:/home/dependencies
    environment:
      - SH_DEPENDENCIES_FILE_PATH=/home/dependencies/
        dependencies.sh
    networks:
      blueborne-net:
        ipv4_address: 10.5.0.5
networks:
  blueborne-net:
    ipam:
      config:
        - subnet: 10.5.0.1/24
\end{lstlisting}
 
It is composed of five services, two of which are the subcomponents of Dockerized Android. The remaining three are briefly described in the following:
\begin{itemize}
    \item \emph{attacker\_phishing}: contains the Gophish component used to craft and send the phishing email that tricks the user into downloading the malicious Fake Spotify app;
    \item \emph{attacker\_web\_server}: contains the webserver used to receive the files sent by the malicious app, which are reverse engineered in order to find information allowing the attacker to exploit the vulnerability on the target device;
    \item \emph{attacker\_blueborne}: the service used by the attacker to execute the attack on the target device and obtain a reverse shell on it.
\end{itemize}

\subsection{Requirements coverage}
In Table~\ref{tab:design-requirements} we have illustrated the defined requirements for the realization of our platform. The following table contains all the requirements and their corresponding status:

\begin{table}[htb!]
    \caption{Requirements evaluation}
    \label{tab:requirements-evaluation}

    \centering
    \begin{tabular}{|c|c|}
        \hline
        \textbf{Requirement ID} & \textbf{Status} \\
        \hline
        F01 & Completed \\
        \hline        
        F02 & Completed \\
        \hline
        F03 & Completed \\
        \hline 
        F04 & Partial \\
        \hline
        F05 & Completed \\
        \hline
        F06 & Partial \\
        \hline
        F07 & Completed \\
        \hline
        F08 & Completed \\
        \hline
        F09 & Completed \\
        \hline
        F10 & Completed \\
        \hline      
    \end{tabular}
\end{table}

Requirement F04, as detailed before, is set to Partial because of the inability to correctly configure all the hardware components (for example the Bluetooth device).
Requirement F06 is set to partial because ADB gives the ability to record the screen out-of-the-box, but this feature was not exposed or made easier to use through the UI.
Finally, requirements F07 (Network Configuration) and F09 (Third-Party Tools integration) are granted by default because of the usage of Docker. The network can be defined in any possible way through the \textit{docker-compose} file, and third-party tools can be easily used together with this system.

\section{Conclusion and Future Developments}
\label{sec:conclusions}
In this work, we have described Dockerized Android, a platform that supports cyber-range designers in realizing mobile virtual scenarios. The application is based on Docker, i.e., a container-based virtualization framework extensively adopted in the cyber-range field for several benefits already mentioned. We described the main components and showed how it is possible to realize a complex cyber kill-chain scenario that involves the utilization of Bluetooth components.   
The architecture has been conceived at the outset as an extensible one. Its feature set can be dynamically enabled or disabled through the docker-compose creator, and some fine-grained options can be configured to customize the scenarios. The strength of this system is its ability to quickly run a mobile component through Docker, with many interesting features out of the box. Furthermore, the centralization of several components increases the overall usability level.  
The cons are all related to compatibility issues with Windows and OS X when running the \textit{Core for Emulator}. While the former will probably be solved with the next updates, the latter is not solvable without significant changes to the OS X implementation. Another limitation is the lack of support for emulating some hardware components, e.g., Bluetooth. For these reasons, the Linux environment as a host machine is strongly recommended. We will also assess the potential benefits of using Dockerized Android in cloud-based environments in future works. Other improvements include the full integration of security-based features in the Android Emulator.
For example, the GPS location could be useful to simulate a realistic route traveled by a simulated user. 
In recent works, cyber ranges are configured by using the high-level SDL (Specification and Description Language) representation \cite{next-generation-cyber-ranges}. Integrating this language in Dockerized Android is relatively easy, as every feature is set through Docker environment variables. Additional efforts will be focused on improving automation features, such as the design of an event-based architecture to simulate complex sequential actions involving human interaction.

\printbibliography

\end{document}